\documentclass[aps,prd,amsmath
,nofootinbib         
,twocolumn         
]{revtex4}
\usepackage{amsmath}
\usepackage{amsfonts}
\usepackage{amssymb}

\allowdisplaybreaks[4]     

\begin{document}

\title{Study of stationary rigidly rotating anisotropic cylindrical fluids with new exact interior solutions of GR. 5. Dust limit and discussion}

\author{M.-N. C\'el\'erier}
\email{marie-noelle.celerier@obspm.fr}
\affiliation{Laboratoire Univers et Th\'eories, Observatoire de Paris, Universit\'e PSL, Universit\'e Paris Cit\'e, CNRS, F-92190 Meudon, France}

\date{12 September 2022}

\begin{abstract}
The present article is the last in a series of five devoted to the study of the effect of anisotropic pressure on the gravitationnal impact of a stationary rigidly rotating cylindrically symmetric fluid with the use of new exact solutions of General Relativity. In the four first papers in this series, three kinds of pressures directed each along one of the principal stresses of the fluid have been considered and new interior solutions of the field equations have been constructed and analyzed. Here, key results and issues raised in these previous works are synthesized and some fundamental notions of general relativity, causality, regularity of Lorentzian manifolds, elementary flatness in the vicinity of a symmetry axis, singularities, physics of angular deficits, weak and strong energy conditions, are revisited. Then, a new derivation of the corresponding dust solution is displayed and shown to correspond indeed to the Lanczos-van Stockum solution shedding new light to this well-known spacetime. 
\end{abstract}


\maketitle 

\section{Introduction} \label{intro}

In general relativity, interior spacetimes generated by a self-graviting source are much less studied than those involving vacuum, since the equations to be solved are much more complicated. However, a number of such solutions are known, but, when the source is a fluid, it is generally a perfect one. The purpose of the present work is the study of the influence of an anisotropic pressure on the interior gravitational properties of a nonperfect fluid. It is customary, in physics, to begin addressing a complex problem with simplifying it. Such an approach has been adopted here. Three types of symmetries have been thus retained: stationarity, axisymmetry, itself specialized to cylindrical symmetry. Then, in the framework of this initial approach, the hypothesis of rigid rotation has been added. To further simplify the field equations and make them integrable, pressure has been decomposed along its three principal stresses: axial, azimuthal and radial. In a series of four articles, here referred to as Paper 1 \cite{C21a}, Paper 2 \cite{C22a}, Paper 3 \cite{C21b} and Paper 4 \cite{C22b}, the effect of each kind of anisotropic pressure on the gravitational impact of a stationary rigidly rotating cylindrically symmetric fluid has been studied with the use of new exact solutions of General Relativity (GR). Three different equations of state have thus been analyzed, one after the other. For two of them, axial (Paper 1 and 2) and azimuthal (Paper 3) pressure, the use of an extra degree of freedom has allowed constructing a general method for integrating the metric equations. It has been then exemplified such as to exhibit particular new exact solutions. Their mathematical and physical properties have been examined and their analyses have led  to dismiss a number of candidates while a fewer ones have been selected as proper solutions. For radial pressure (Paper 4), the degree of freedom disappears in favor of a splitting into three classes of solutions each defined by a differential equation issued from the factorisation of the field equations. In Paper 4, only one class has been integrated, the other two remaining in the form of simplified differential equations. However, despite the assumptions made for mathematical purpose, the properties of the generated spacetimes are such as to enable proposals for actual physical applications.

As it is well-known, there exist genuine solutions of Einstein's field equations which cannot be used to characterize a Lorentzian manifold. To deal with this issue happens to be one of the main challenges of this series of works. The subject of the present article, named Paper 5 in the following, is to come back on the different aspects of the encountered difficulties and thus to go deeper into some fundamental issues frequently met in GR.

The most common and most severe pathology is that of a possible non-Lorentzian metric signature. It is linked to the possible occurrence of closed timelike curves which seems to be rather generically present in spacetimes sourced by rapidly rotating infinite cylinders \cite{M66,T74}. It is considered here in Sec. \ref{sign} by referring to results exhibited in Papers 1-4.

 In contrast, axisymmetry is a rather well-understood notion whose definition is recalled in Sec. \ref{regul} and whose characterizing equation is systematically applied to the studied spacetimes. However, another notion that needs some clarifications is that of regularity and, in particular, the regularity of the symmetry axis, here specialized to the cylindrically symmetric case. Moreover, the other key requirement of elementary flatness is indeed a topic too broad to be definitively settled in the present article, but a related discussion and a number of clues are given in Sec. \ref{regul}.

Finally, it must be stressed that the dust limit cannot be recovered by merely setting the pressure to vanish in the different solutions displayed in Papers 1-4. Remember that a key auxiliary function introduced there for the calculations is $h$, the ratio of the non-zero component of the pressure over the energy density. The absence of such dust limit is due to the fact that all the calculations conducted to obtain these solutions have been made under the assumption $h \neq 0$. However, it is shown in Sec. \ref{rrd} that the long-known Lanczos-van Stockum spacetimes for cylindrically symmetric dust \cite{L24,vS37} can be recovered as solutions of the field equations of Papers 1-4 written with vanishing pressure components. A new construction of this dust solution, using the method followed in Papers 1-4 is here proposed. It allows to follow step by step the simplification of this solution from a five parameter crude mathematical solution to a one parameter final one. Moreover, a number of its physical properties are analyzed from a new point of view in this section.

Sec. \ref{concl} is devoted to the conclusion.

\section{Metric signature and causality violation} \label{sign}

As recalled in the Introduction, many actual solutions of Einstein's equations cannot be used as proper metrics for GR manifolds since they are not Lorentzian. For cylindrically symmetric spacetimes, the problem appears rather generically, due to the extended use of polar coordinates. This is the reason why this feature has been carefully examined in turn for each solution in the series considered in Papers 1-4. In general, these analyses have led to one of the following conclusions. Either demanding the metric signature to be Lorentzian resulted in some constraints on the parameters and/or in limits on the coordinates of the solutions. In this case, the solutions could be retained provided these constraints applied. Or the metric signature was forced to remain non-Lorentzian and the solutions have been dismissed.

This has been indeed the case for Class 2 (i) of Paper 3 which has been ruled out owing to an improper sign for the metric function $l$. Now, what about $l$ becoming, say, negative, while the three other metric functions $f$, $k$ and $\textrm{e}^\mu$ are positive definite? Given the form of the line element recalled as (\ref{metric}) in Sec. \ref{rrd}, $l$ appears as being the $g_{\phi \phi}$ coefficient of the metric which, being negative, implies causality violation. Indeed, the lines $\{t=const., r=const., z=const.\}$ are thus closed timelike curves and can be drawn from and to any event in spacetime. Therefore, solutions belonging to Class 2 (i) despite having been dismissed in Paper 3 could be considered of interest for applications where such ''pathologies'' are not considered as too strong a drawback. Actually, closed timelike lines appear in other GR solutions such as the G\"odel and the Kerr metrics, but in both cases, whether the sources are made of reasonable matter is unknown. In the present case of a well-defined non perfect fluid, it had seemed reasonable to avoid such issues by choosing to rule out the corresponding spacetimes in Paper 3, but this might not necessarily be the last word of it.

The other classes of solutions displayed in Papers 1-4 have been constructed such as to exhibit a proper Lorentzian signature. Hence closed timelike curves are not present in those spacetimes. However, such a demand has often implied constraints on the solutions themselves. In most cases, i.e., for Paper 1 case (i) and (ii), for Paper 2 Class A and for Paper 4 Class I, the constraint results in limits on the allowed values of the $h$ ratio, transforming to explicit or implicit limits on the $r$ coordinate, hence, on the radius of the cylinder. An analogous constraint has already been known for long in the dust case, where the radial coordinate is bounded by van Stockum's $1/a$ parameter \cite{vS37}. It is therefore interesting to note that an anisotropic pressure does not change the nature of the constraints imposed on the geometry of the source by the metric signature, even though it changes their magnitudes.

The properties of the azimuthally directed pressure fluids of Paper 3 are somehow different. Here, not only the imposed constraints have as an effect to put limits on the size of the cylinder through explicit or implicit limits on the $h$ ratio, but it has another consequence impacting the value $h_0$ of this ratio on the symmetry axis which appears to be also bounded. In this configuration, the relative magnitude of the pressure over the energy density on the axis cannot be any. Indeed, for Class 1 where the strong energy condition is satisfied, $h_0$ must verify $h_0 > 0.193419$, and for Class 2 (ii), where only the weak energy condition holds, $h_0$ is limited by $-1<h_0<-0.42$. This emphasizes the special status of this kind of anisotropy as regards the gravitational effect of pressureless and axially or radially directed pressure fluids. Indeed, since the value of $h_0$ is linked to that of the amplitude of the angular velocity of the fluid on the axis, by (84) for Class 1 and (156) for Class 2 in Paper 3, this angular velocity is constrained for each class in a way independent on the magnitude of the pressure itself. Indeed, even though the amplitudes of the angular velocity and of the ratio of pressure over energy density on the axis are related, their allowed range is fixed whatever the particular values selected by the physical system inside this range.

The above analysis shows that fundamentally different properties are exhibited by the spacetimes and their sourcing fluids depending on the way pressure occurs and is directed.

\section{Axisymmetry, regularity and elementary flatness} \label{regul}

In Papers 1-4, the examined solutions have been confronted to three particular requirements: axisymmetry, regularity and elementary flatness.

Axial symmetry is characterized by an isometry generated
by a spacelike Killing vector, say $\vec{\xi}$, whose orbits are closed compact curves. The axis is the set of fixed points that are unchanged by this isometry. The usual approach, that has been adopted in this series of articles, consists in choosing a coordinate, say $\phi$, adapted to the Killing vector which generates the isometry so that $\vec{\xi}= \partial_{\phi}$. The advantage is that the metric and other geometrical and physical objects are independent of this coordinate. However, it has been advocated, e. g., by Carot \cite{C00}, that a chart with such a choice of coordinate cannot contain points belonging to the symmetry axis. Anyhow, it is necessary to include the points on the axis into the underlying manifold, otherwise the spacetimes would be geodesically incomplete and a geodesically incomplete spacetime should be considered as singular. This might be done through the use of a coordinate chart covering the symmetry axis and related to the former by a diffeomorphism in the overlap region. Such a chart usually exists and is Cartesian-like. However, for mathematical simplicity, the polar coordinates have been kept here to integrate the field equations.
 
Hence, to ensure the existence of a proper symmetry axis, the usual axisymmetry condition \cite{W96,D06} has been imposed on every candidate solution without exception, since it is an unavoidable feature of cylindrical symmetry. It is recalled here as
\begin{equation}
l \stackrel{0}{=} 0, \label{r0a}
\end{equation}
where $\stackrel{0}{=}$ notifies that the values are to be taken at the limit on the axis. Therefore every exact solution displayed in this series of articles, i.e., Papers 1-5, has been required to fulfill this constraint.

To be considered as a smooth Lorentzian manifold, able to represent an appropriate GR spacetime, the solutions need to be non singular, save perhaps at some non-essential locations. Now, an essential location in cylindrical symmetry is the symmetry axis which has been specially examined from this point of view for each class of solutions. However, some other hypersurfaces living on the new spacetimes have been identified as possibly singular. Fortunately, most of them are defined by limiting values of the $h$ ratio and therefore easy to remove without generating geodesic incompleteness. 

Now, a method for identifying the existence of hidden singularities on the axis has been proposed by Mars and Senovilla \cite{M93,C00,S09}. It is based on the requirement of elementary flatness which is a key feature of GR. It consists in checking whether the ratio of the circumference over the radius of an infinitesimally small circle around the axis departs or not from the value $2 \pi$. The circle is defined as the orbit of the spacelike Killing vector $\vec{\xi}$ generating the isometry. This Killing vector must therefore satisfy the so-called ''regularity'' condition \cite{S09}
\begin{equation}
\frac{\partial_{\alpha}\left(\xi^{\mu} \xi_{\mu}\right) \partial ^{\alpha}\left(\xi^{\nu} \xi_{\nu}\right)}{4 \xi^{\lambda}\xi_{\lambda}} \stackrel{0}{=} 1, \label{r0b}
\end{equation}
which, for cylindrical symmetry and with the coordinate frame retained here becomes below (\ref{r42}). 

Now, as it has been claimed by Wilson and Clarke \cite{W96}, this so-called ''regularity'' condition \cite{S09}, which is actually an elementary flatness condition, does not ensure the smoothness of the manifold on the axis. This statement can find a confirmation by the analysis of the different solutions displayed in Papers 1-5 that follows.

In order to select solutions that could be used for astronomical or cosmological purpose, the axisymmetry and ''regularity'' conditions have systematically been considered and confronted to other standard methods for exhibiting singularities. The results are as follows.

A first set of solutions have been shown to satisfy both axisymmetry and ''regularity'' conditions, while exhibiting no metric singularities wherever: both Class 1 and 2 of Paper 3, representing solutions for a fluid with azimuthally directed pressure and, of course, the Lanczos-van Stockum dust solution analyzed here in Sec. \ref{rrd}.

The axially directed pressure solution displayed in Paper 1, and its subcases (i) and (ii) satisfies a large number of conditions: axisymmetry, ''regularity'', the increased elementary flatness conditions of Debbasch et al. \cite{D06}, while its Kretschmann scalar diverges for $h=-1$ and $h=0$. However, since these two values correspond to the limits of the allowed interval for this ratio, the corresponding points can be excluded without damage from the spacetime. More unexpected, this curvature invariant diverges also on the axis, although all the ''regularity'' constraints are fulfilled by the solution. This example shows that ''regularity'' and elementary flatness conditions are not sufficient to ensure the smoothness of cylindrical spacetimes on their symmetry axis.

Conversely, two classes of solutions fulfilling the axisymmetry but NOT the ''regularity'' conditions, while exhibiting a finite Kretschmann scalar on the axis, have been found: Class A of Paper 2, for axial pressure, and Class I of Paper 4, for radial pressure. Indeed, these spacetimes suffer from an angular deficit due to their non-satisfying the ''regularity'' condition \cite{K14}. This criterion of elementary flatness actually tests the local Riemannian (here Lorentzian) character of a manifold. The problematic region of the axis represents a singularity in the sense that if it is excluded from the spacetime, then the resulting manifold is geodesically incomplete \cite{S85}. However, an angular deficit is a characteristic feature of a distributional source, such as a topological defect or a cosmic string \cite{G87}. Indeed, an infinitely thin object, when glued in the place of the removed 2-dimensional surface of the axis might therefore constitute an adequate source for both classes of spacetimes.

The solution displayed in Appendix B of Paper 2 is still more interesting. While it is well-behaved when the ''regularity'' condition is not applied, it becomes non-Lorentzian, i.e., its $l$ metric function changes sign when this condition is imposed. This change of signature, where $g_{\phi \phi}$ gains the sign of $g_{tt}$, implies the occurrence of closed timelike curves inside the cylinder, which has been discussed in Sec. \ref{sign}.

The above summary of the regularity features of the solutions studied in the framework of this work devoted to the unravelling of the properties of pressure anisotropy in cylindrical symmetry shows that the concept of regularity is far from being well understood when dealing with polar coordinates and axial symmetry. Therefore, the exhibited features might display some clues to help improving our approach of the issue.

\section{Energy conditions} \label{ec}

Other important constraints have been considered when examining the different classes of solutions displayed in Papers 1-4. To be sure that the fluid sourcing the spacetimes is physically well-behaved, the weak energy condition, $\rho\geq 0$, has been systematically imposed to every solution adopted.

However, since they might be referred to known systems such as dark energy, negative pressure solutions have been accepted. Such are case (ii) described in Paper 1, Class 2 (ii) from Paper 3 and Class I from Paper 4. Note moreover that negative energy-density contributions may turn out to be essential for a satisfactory defect solution in nonsimply connected manifolds \cite{K14,G75}. Since a cylindrically symmetric spacetime with a removed axis is nonsimply connected, such a property might be of use in a defect framework. Finally, all the other solutions exhibit a positive pressure which is an ordinary feature of ''standard'' fluids.

Now, once implemented into the different solutions, all the constraints discussed in Secs. \ref{sign} and \ref{regul} and in the present section lead to reducing the initial number of integration constants to a few independent parameters. Save for the radially directed pressure solutions of Class I which finally depends on two independent parameters $h_0$ and $c$ (or equivalently the amplitude of the rotation scalar on the axis $\omega_0$), all the other solutions displayed in this work end up as one parameter solutions. This parameter can be chosen to be either $h_0$ or $\omega_0$, since both are dependent one from the other. This is the same number as for the dust solution considered in Sec. \ref{rrd}, but much less than in the case of the four-parameter vacuum solution of Lewis-Weyl \cite{L32,W19}. It is therefore interesting to note that the drop from four to one parameter of the metric is quite generally identical when going from the exterior to the interior spacetime of a stationary rigidly rotating matter cylinder whatever the equation of state of the gravitating fluid.
 
\section{Rigidly rotating dust. Lanczos-van Stockum solution revisited.} \label{rrd}

The series of anisotropic pressure solutions displayed in Papers 1--4 seem, at first glance, to have no dust limit since setting $h=0$ directly into the expressions of these solutions does not allow to recover the known form of the metric of the corresponding cylindrical dust spacetime. Now, $h \neq 0$ is a key property of the pressure over energy density amplitude ratio used to derive these anisotropic solutions. This is the reason why trying to make it vanish leads to an inaccurate result. However, interior spacetimes sourced by a stationary rigidly rotating dust cylinder, first exhibited by Lanczos in 1924 \cite{L24}, then rediscovered by van Stockum in 1937 \cite{vS37}, can be derived using a method analogous to that implemented in the pressure cases formerly studied in the framework of the present series. The van Stockum form is therefore recovered (for vanishing cosmological constant) while some interesting features are exhibited in the process.

\subsection{Setting the main tools} \label{sp}

General properties of interior spacetimes sourced by general stationary fluids have been analyzed by C\'el\'erier and Santos \cite{CS20}. Since part of the equations displayed in that paper will be of use in the present article, they are recalled below and specialized to rigidly rotating dust.

The source of the here considered spacetime is a cylindrically symmetric zero-pressure fluid in  stationary motion, rigidly rotating around its axis. It is dissipative and bounded by a cylindrical surface $\Sigma$. Its stress-energy tensor, whose general expression is given by (1) of C\'el\'erier and Santos \cite{CS20}, can therefore be written under the form
\begin{equation}
T_{\alpha \beta} = \rho  V_{\alpha}V_{\beta}, \label{r1}
\end{equation}
where $\rho$ denotes the energy density of the fluid and $V_\alpha$, its timelike 4-velocity  satisfying
\begin{equation}
V^\alpha V_\alpha = -1. \label{r2}
\end{equation} 
As for the previous spacetimes of this series a spacelike hypersurface orthogonal Killing vector $\partial_z$ is assumed, in order to facilitate its possible subsequent junction to an exterior Lewis metric. Therefore, in geometric units $c=G=1$, the line element reads
\begin{equation}
\textrm{d}s^2=-f \textrm{d}t^2 + 2 k \textrm{d}t \textrm{d}\phi +\textrm{e}^\mu (\textrm{d}r^2 +\textrm{d}z^2) + l \textrm{d}\phi^2, \label{metric}
\end{equation}
$f$, $k$, $\mu$, and $l$ being real functions of the radial coordinate $r$ only, such as to account for stationarity. The cylindrical feature forces the coordinates to conform to the following ranges:
\begin{equation}
-\infty \leq t \leq +\infty, \quad 0 \leq r < +\infty, \quad -\infty < z < +\infty, \quad 0 \leq \phi \leq 2 \pi, \label{ranges}
\end{equation}
the two limits of the coordinate $\phi$ being topologically identified. These coordinates are denoted $x^0=t$, $x^1=r$, $x^2=z$, and $x^3=\phi$.

Rigid rotation allows the choice of a frame corotating with the fluid \cite{D06,C21a,CS20}. Thus, its 4-velocity can be written as
\begin{equation}
V^\alpha = v \delta^\alpha_0, \label{r3}
\end{equation}
with $v$ a function of $r$ only. Therefore, the timelike condition for $V^\alpha$ displayed in (\ref{r2}) becomes
\begin{equation}
fv^2 = 1. \label{timeliker}
\end{equation}
As for the anisotropic pressure cases, an auxiliary $D(r)$ function, already used by van Stockum who defined it directly to be the radial $r$ coordinate, is here defined by
\begin{equation}
D^2 = fl + k^2. \label{D2}
\end{equation}

\subsection{Field equations} \label{fe}

Using (\ref{r3}) into (\ref{r1}), the components of the stress-energy tensor matching the five nonvanishing components of the Einstein tensor are obtained, and the five corresponding field equations can be written as

\begin{equation}
G_{00} = \frac{\textrm{e}^{-\mu}}{2} \left[-f\mu'' - 2f\frac{D''}{D} + f'' - f'\frac{D'}{D} + \frac{3f(f' l' + k'^2)}{2D^2}\right]= \kappa\rho f, \label{G00}
\end{equation}
\begin{equation}
G_{03} =  \frac{\textrm{e}^{-\mu}}{2} \left[k\mu'' + 2 k \frac{D''}{D} -k'' + k'\frac{D'}{D} - \frac{3k(f' l' + k'^2)}{2D^2}\right] = - \kappa\rho k, \label{G03}
\end{equation}
\begin{equation} 
G_{11} = \frac{\mu' D'}{2D} + \frac{f' l' + k'^2}{4D^2} = 0, \label{G11}
\end{equation}
\begin{equation}
G_{22} = \frac{D''}{D} -\frac{\mu' D'}{2D} - \frac{f'l' + k'^2}{4D^2} = 0, \label{G22}
\end{equation}
\begin{equation}
G_{33} =  \frac{\textrm{e}^{-\mu}}{2} \left[l\mu'' + 2l\frac{D''}{D} - l'' + l'\frac{D'}{D} - \frac{3l(f' l' + k'^2)}{2D^2}\right] =  \kappa \rho\frac{ k^2}{f}, \label{G33}
\end{equation}
where the primes denote differentiation with respect to $r$.

Contrary to what happened in the cases studied in previous Papers 1--4, the number of independent differential equations, i.e., five, is exactly the same as that of the unknown functions, the four metric functions and the energy density. Therefore the problem is perfectly closed and no degree of freedom is left pending.

\subsection{Conservation of the stress-energy tensor} \label{bi}

The conservation of the stress-energy tensor is implemented by the Bianchi identity, whose general form is available as (56)-(57) of C\'el\'erier and Santos \cite{CS20}. Specialized to the present case, it becomes
\begin{equation}
T^\beta_{1;\beta} = \rho \frac{f'}{2f} = 0. \label{Bianchi}
\end{equation}

\subsection{Useful reminder}

Two important equations established in previous works and which will be needed for the following calculations are briefly recalled below. The first one, initially displayed as (14) in Debbasch et al. \cite{D06}, is written here as
\begin{equation}
kf' - fk' = -2c D, \label{r4}
\end{equation}
where $c$ is an integration constant and the factor -2 is added for further convenience. This first-order ordinary differential equation in $k$ possesses, as a solution, see (24) of Paper 1 where the minus sign of (\ref{r4}) has been implemented,
\begin{equation}
k = f \left(c_k + 2c \int^r_0 \frac{D(v)}{f^2(v) } \textrm{d} v \right). \label{r5}
\end{equation}

\subsection{Solving the field equations} \label{sol}

The Bianchi identity (\ref{Bianchi}) implies
\begin{equation}
f'= 0, \label{r6}
\end{equation}
which can be integrated as
\begin{equation}
f= c_f, \label{r7}
\end{equation}
where $c_f$ is an integration constant.

Now, adding (\ref{G11}) and (\ref{G22}), one obtains
\begin{equation}
D''= 0, \label{r8}
\end{equation}
that, according to Paper 3, can be integrated as
\begin{equation}
D= r + c_2. \label{r9}
\end{equation}
A useful calculation tool will be
\begin{equation}
\frac{D'}{D}= \frac{1}{r + c_2}. \label{r9a}
\end{equation}

Then, (\ref{r7}) and (\ref{r9}) inserted into (\ref{r5}) give
\begin{equation}
k = c_f \left(c_k + 2c \int^r_0 \frac{v + c_2}{c_f^2 } \textrm{d} v \right), \label{r10}
\end{equation}
which can be integrated as
\begin{equation}
k = c_f c_k + \frac{c r}{c_f}(r + 2 c_2). \label{r11}
\end{equation}

As usual, $l$ follows from the definition (\ref{D2}) of the auxiliary function $D$. It reads
\begin{equation}
l = \frac{(r+c_2)^2}{c_f} - \frac{1}{c_f} \left[c_f c_k + \frac{cr(r+2c_2)}{c_f}\right]^2. \label{r12}
\end{equation}

Now, (\ref{r11}) differentiated with respect to $r$ is inserted into (\ref{G11}) together with (\ref{r6}), (\ref{r9}) and its derivative $D'$ such as to obtain
\begin{equation}
\mu' = - \frac{2 c ^2}{c_f^2}(r + c_2), \label{r13}
\end{equation}
which can be integrated as
\begin{equation}
\textrm{e}^{\mu} = c_{\mu} \exp\left[- \frac{c ^2}{c_f^2}r(r + 2c_2)\right], \label{r14}
\end{equation}
where $c_{\mu}$ is an integration constant that can be eliminated by a proper rescaling of the $r$ and $z$ coordinates as in Papers 2--4.

Then, (\ref{r13}) differentiated with respect to $r$ is inserted into (\ref{G00}) together with (\ref{G11}) and the expressions for $f$, $D$ and derivatives so that to yield
\begin{equation}
\rho \textrm{e}^{\mu} = \frac{4 c ^2}{\kappa c_f^2}, \label{r15}
\end{equation}
that gives, with (\ref{r14}) inserted
\begin{equation}
\rho = \frac{4 c ^2}{\kappa c_f^2}\exp\left[\frac{c ^2}{c_f^2}r(r + 2c_2)\right]. \label{r16}
\end{equation}

\subsection{Axisymmetry condition} \label{axi}

The axisymmetry condition which reads \cite{S09}
\begin{equation}
l \stackrel{0}{=} 0, \label{r17}
\end{equation}
becomes, with (\ref{r12}) inserted,
\begin{equation}
c_2^2 = c_f^2 c_k^2. \label{r18}
\end{equation}
Taking the square root of the above, it comes
\begin{equation}
c_k = \epsilon \frac{c_2}{c_f}, \label{r19}
\end{equation}
where $\epsilon = \pm 1$.

Implementing constraint (\ref{r19}) into the expressions for the metric functions displayed in Section \ref{sol}, one obtains
\begin{equation}
f= c_f, \label{r22}
\end{equation}
\begin{equation}
\textrm{e}^{\mu} = \exp\left[- \frac{c^2}{c_f^2} r(r + 2c_2)\right], \label{r23}
\end{equation}
\begin{equation}
k = \epsilon c_2 + \frac{c}{c_f} r(r+ 2c_2), \label{r24}
\end{equation}
\begin{equation}
l = \frac{(r+c_2)^2}{c_f} - \frac{1}{c_f} \left[\epsilon c_2 + \frac{cr(r+2c_2)}{c_f}\right]^2, \label{r25}
\end{equation}
\begin{equation}
\rho = \frac{4 c ^2}{\kappa c_f^2}\exp\left[\frac{c ^2}{c_f^2}r(r + 2c_2)\right]. \label{r26}
\end{equation}

\subsection{Constraints from the metric signature} \label{msign}

Owing to the form (\ref{metric}) of the metric, a Lorentzian signature is obtained provided the metric functions be either all positive or all negative definite.  

Now, at the axis, where $r=0$, the functions displayed in Sec. \ref{axi} take the values
\begin{equation}
f \stackrel{0}{=} c_f, \label{r27}
\end{equation}
\begin{equation}
\textrm{e}^{\mu} \stackrel{0}{=} 1, \label{r28}
\end{equation}
\begin{equation}
k \stackrel{0}{=} \epsilon c_2, \label{r29}
\end{equation}
\begin{equation}
l \stackrel{0}{=} 0, \label{r30}
\end{equation}
\begin{equation}
\rho \stackrel{0}{=} \frac{4 c ^2}{\kappa c_f^2}, \label{r31}
\end{equation}
where $\stackrel{0}{=}$ denotes that the values are to be taken at the axis. Since the sign exhibited by a metric function at the limit at the axis must be kept the same in the whole spacetime, the values listed above impose some constraints on the parameters. Indeed, $\textrm{e}^{\mu}$, being positive there, is bound to remain such everywhere which is actually forced upon it by the exponential in (\ref{r23}). Therefore, all the other metric functions must be positive and the following constraints emerge from the values on the axis:
\begin{equation}
c_f > 0, \label{r32}
\end{equation}
\begin{equation}
\epsilon c_2 > 0. \label{r33}
\end{equation}

Now it must be ensured that positiveness is indeed kept by the four metric functions about the whole spacetime. This is obvious for $f$ and $\textrm{e}^{\mu}$. It must be examined more thoroughly for $k$ and $l$.

The positiveness of the metric function $k$, given by (\ref{r24}) induces
\begin{equation}
\left(r+c_2+ \sqrt{c_2^2-\frac{\epsilon c_2 c_f}{c}}\right) \left(r+c_2- \sqrt{c_2^2-\frac{\epsilon c_2 c_f}{c}}\right) > 0. \label{r34}
\end{equation}
The analysis of the constraints inferred by this inequality yields
two possible cases.

(a) $c_2=0$ and $c>0$.

(b) $c_2<0$, that implies $\epsilon=-1$ from (\ref{r33}), $r \leq -c_2$ and $c_f/c<-c_2$.

Now, the consequences of imposing $l \geq 0$ in each of the two above cases (a) and (b) are considered in turn.

\hfill

(a) Inserting $c_2=0$ into (\ref{r25}) gives
\begin{equation}
l = \frac{r^2}{c_f}\left(1 - \frac{c^2}{c_f^2}r^2 \right). \label{r35}
\end{equation}
This expression is $\geq 0$ provided $r\leq c_f/c$, which is coherent since $c$ takes positive values in this case and $c_f$ is always positive from (\ref{r32}). One recognizes easily in this case, the solution displayed, after Lanczos, by van Stockum (vS) with the following correspondence for the notations:

$c/c_f \rightarrow$ $a$ (vS)

 $c_f \rightarrow $ $1$ (vS)

\hfill

(b) $\epsilon=-1$ inserted into (\ref{r25}) allows to write $l \geq 0$ as
\begin{equation}
r\left(c r + c_f + 2 c c_2\right) \left(r-\frac{c_f}{c}\right) \left(r +2c_2\right) \geq 0. \label{r36}
\end{equation}
A straightforward examination of the different combinations of the signs of these products shows that none of them leads to a mathematically or physically well-behaved result. This (b) case is therefore to be dismissed and only the Lanczos-van Stockum-like case (a) is elligible as a genuine interior solution for a stationary rigidly rotating dust cylinder.

\hfill

This result shows that the well-known dust solution can be indeed recovered as the zero-pressure limit of the anisotropic pressure series of solutions. It is given below as a reminder for further use, using the vS notation $a \equiv c/c_f$, but keeping $c_f$ unspecified for the moment. It thus reads
\begin{equation}
f= c_f, \label{r37}
\end{equation}
\begin{equation}
\textrm{e}^{\mu} = \textrm{e}^{- a^2 r^2}, \label{r38}
\end{equation}
\begin{equation}
k = a r^2, \label{r39}
\end{equation}
\begin{equation}
l = \frac{1}{c_f} r^2(1- a^2 r^2), \label{r40}
\end{equation}
\begin{equation}
\rho = \frac{4 a ^2}{\kappa}\textrm{e}^{a^2 r^2}, \label{r41}
\end{equation}
with a limiting value for the radial coordinate given by $r<1/a$ so that the metric signature be well-behaved.

\subsection{The regularity condition} \label{reg}

The regularity condition has been discussed in Sec. \ref{regul}. It is however interesting to see its effect on metric (\ref{r37})--(\ref{r40}). It can indeed be written \cite{S09,D06}
\begin{equation}
\frac{\textrm{e}^{-\mu}l'^2}{4l} \stackrel{0}{=} 1. \label{r42}
\end{equation}
Differentiating (\ref{r40}) with respect to $r$ and inserting it into the left-hand side of (\ref{r42}) together with (\ref{r38}) and (\ref{r40}), one obtains
\begin{equation}
\frac{\textrm{e}^{-\mu}l'^2}{4l} = \textrm{e}^{a^2 r^2} \frac{(1- 2 a^2 r^2)^2}{c_f(1- a^2 r^2)}, \label{r43}
\end{equation}
which, evaluated at the axis, gives
\begin{equation}
\frac{1}{c_f} = 1. \label{r44}
\end{equation}
The regularity condition is therefore verified by this solution,  provided $c_f=1$. In this case, the solution takes exactly the van Stockum form, which appears therefore indeed as the dust limit of the problem.

It is interesting to notice that, $c_f=1$, which was merely assumed by van Stockum, is here given a justification through the requirement of elementary flatness, i.e., the absence of any angular deficit in the vicinity of the axis.

\subsection{Physical properties} \label{phy}

Many physical properties of the Lanczos-van Stockum (LvS) solution have been displayed in the literature \cite{T74,B80}. However, some points have been less thoroughly considered. This is the reason why they are examined here.

The hydrodynamical properties of the dust fluid can be calculated using the former results obtained by C\'el\'erier and Santos \cite{CS20} in the most general case with pressure, specialized to the present case. In the rigid rotation case, the usually non-zero component of the acceleration vector can be written as
\begin{equation}
\dot{V}_1 = -\Phi = \frac{f'}{2f}. \label{r45}
\end{equation}
For the dust solution, $f'$ and therefore $\dot{V}_1$ vanish. The dust fluid experiment no acceleration inside the cylinder.

Two non-zero components have been identified by C\'el\'erier and Santos \cite{CS20} for the rotation tensor in the pressure case. Here they become
\begin{equation}
2 \omega_{01} = -(fv)' = 0, \label{r46}
\end{equation}
owing to the timelike condition (\ref{timeliker}) and the vanishing of the $f$ derivative with respect to $r$. The actual non-vanishing component of the rotation tensor follows as
\begin{equation}
2 \omega_{13} = -(kv)'. \label{r46}
\end{equation}
Using the derivative of $k$ with respect to $r$ and the timelike condition (\ref{timeliker}), one obtains
\begin{equation}
2 \omega_{13} = -2ar. \label{r46}
\end{equation}
Finally, the rotation scalar is given by
\begin{equation}
\omega^2 = \frac{f}{4 \textrm{e}^{\mu} D^2 f^2 v^2}\left(-f k' v^2 \right)^2, \label{r46}
\end{equation}
which becomes, with the metric functions and the timelike condition inserted,
\begin{equation}
\omega^2 = a^2 \textrm{e}^{a^2 r^2}. \label{r47}
\end{equation}
On the axis, where $r=0$, the amplitude of this scalar takes the value $a$, that is the van Stockum notation for parameter $c$. It is therefore interesting to remark that the property proven in Appendix B of Paper 2 for the pressure case, and stating that the amplitude of the rotation scalar at the axis is given by the parameter $c$, still holds for dust. Hence, any one parameter interior solution of GR sourced by a stationary rigidly rotating cylinder of fluid, whatever its equation of state, appears to be fully determined by the angular velocity on the axis. This feature has been already stressed by Bonnor \cite{B80} for the dust case, but its generalisation to the anisotropic cases is a new result of the present series of papers.

\section{Conclusion} \label{concl}

The results displayed in the series of former articles devoted to the study of the influence of anisotropic pressure on interior spacetimes sourced by stationary rigidly rotating cylindrically symmetric fluids have been here summarized and discussed. Some important features have been exhibited such as the implications of the imposed  Lorentzian nature of the metric signature, in order to prevent the occurrence of closed timelike curves, on the extent of the allowed limits of the solution parameter, depending on the pressure configuration.

The status of the so-called ''regularity'' condition, which had already been discussed by Wilson and Clarke \cite{W96}, has been provided with new enlightenment through its application to the new solutions and the issuing correspondence with other regularity and elementary flatness tracers.

Added to the different ways of applying the energy conditions recalled here, these constraints have generated one parameter spacetimes, save for the case of the radial pressure Class I where the solutions depend on two parameters. Hence, this property, which has already been stressed by Bonnor \cite{B80} for the dust case, seems to be generalizable to nearly the whole set of anisotropic pressure cases. It can be contrasted to the four parameter Lewis-Weyl exterior solution known as its vacuum counterpart.

Another key contribution of the present paper is, after explaining why the dust limit of the anisotropic pressure configurations cannot be recovered by merely setting to vanish the auxiliary function $h(r)$ used for the integration of the latter solutions, to have shown that this dust limit can be obtained by a method strictly analogous to the one employed to derive the anisotropic pressure solutions. Displaying thus a new detailed derivation of the LvS solution has therefore allowed to dig deeper into the assumptions formerly made and to identify the effects of the different constraints imposed at each step.

Finally, it is important to stress that, in the process of selecting, among the mathematically well-behaved solutions, those that could be physically acceptable, a more or less subjective choice has to be made. This is where both general methods displayed in Papers 2 and 3 for integrating other solutions come in. They give the interested reader a recipe to generate other new solutions adapted to some given physical configuration. Of course, the field equations of GR being nonlinear, the use of particular solutions of the set to generate more sophisticate ones must be handled very carefully. Anyhow, each nonzero pressure component of the series can be considered as a mere approximation of a more physically complete spacetime. Actually, for any application where a principal stress component of the pressure dominates, a perturbation approach around the corresponding solution can be of use. Now, part of the solutions described can be applied to basically ''standard'' fluids. Even when they exhibit a negative pressure can they be related to metastable states (see the Conclusion of Paper 4).  However, a quasi-infinite cylinder of matter might also represent some kind of topological defects or cosmic(super)-strings. For such applications, new solutions with different properties might be needed. As an example, a characteristic feature of topological defects is their angle-deficit which, in this case, is a required property aiming at a non-satisfaction of the ''singularity'' condition. Hence the interest of such a series of very widely open results.

\end{document}